\documentclass[a4paper,12pt]{article}

\usepackage{latexsym,amsmath,amsfonts,amssymb}
\usepackage{multirow}
\usepackage[dvips]{graphicx}
\usepackage{mathrsfs,mathtools}

\newcommand{\sect}[1]{\setcounter{equation}{0}\section{#1}}

\def\beqa{\begin{eqnarray}}
\def\eeqa{\end{eqnarray}}
\def\beq{\begin{equation}}
\def\eeq{\end{equation}}

\newcommand{\be}{\begin{equation}}
\newcommand{\ee}{\end{equation}}

\begin{document}

\begin{titlepage}

\begin{flushright}
{QMUL-PH-09-03}
\end{flushright}

\vspace{0.6cm}

\begin{center}
  {\Large \bf Null Wilson loops with a self-crossing and the Wilson
      loop/amplitude conjecture}

\vskip 0.8cm

{\bf George Georgiou}\\
{\sl
Centre for Research in String Theory \\ Department of Physics\\
Queen Mary, University of London\\
E1 4NS, United Kingdom}\\
g.georgiou@qmul.ac.uk
\vskip 1.2cm

\end{center}

\begin{abstract}
The present study illuminates the relation between null
  cusped Wilson loops and their corresponding amplitudes. We find
  that, compared to the case with no self-crossing, the one loop
  expectation value of a self-intersecting Wilson loop develops an
  additional $1/\epsilon$ singularity associated to the intersection.
  Interestingly, the same $1/\epsilon$ pole exists in the finite
  part of the one loop amplitude, appearing in the BDS conjecture, at
  the corresponding kinematic limit.
  At two loops, we explore the behaviour of the remainder function
  $R$, encoding the deviation of the amplitude from the BDS
  conjecture. By analysing the renormalisation group equations for the
  Wilson loop with a simple self-crossing, we argue that, when
  approaching the configuration with a self-crossing ($u_2\rightarrow
  1$, $u_1\approx u_3$), $R$ diverges in the imaginary direction like
  $R\sim i \pi \log^3(1-u_2)$.  This behaviour can be attributed to
  the non-trivial analytic continuation needed when passing from the
  Euclidean to the physical region and suggests that $R$ has a branch
  cut in the negative $u_2$ axis when the two other cross ratios are
  approximately equal ($u_1\approx u_3$).
\end{abstract}

\vfill

\end{titlepage}

\sect{Introduction}\label{Intro}

One of the most important classes of
quantities calculated in quantum field theory is that of the on-shell
scattering amplitudes. In fact, the knowledge of the scattering matrix
completely specifies the theory, perturbatively.  Moreover, scattering
amplitudes of gauge theories frequently exhibit structures and
symmetries which are not at all apparent from the Lagrangian
formulation of the system.

 One such example is the maximally helicity violating (MHV) tree level amplitudes of an arbitrary number of gluons
\cite{PT}. Another example is the iterative structure of the multi-loop MHV scattering amplitudes in the planar maximally super-symmetric Yang-Mills theory (${\cal N}=4$ SYM). Such a structure was first observed in \cite{Anastasiou:2003kj}
for the case of the four gluon planar amplitude at two loops. In the same parer, it was conjectured that this iterative structure may also hold for an arbitrary number of gluons.

Subsequently, an explicit calculation of the four gluon amplitude at three loops led the authors of \cite{Bern:2005iz} to propose a
conjecture for the all-loop expression of the n-point MHV amplitude. From the work of \cite{IRexp} , it is known that the soft and collinear singularities of any gauge theory amplitude exponentiate.
What is remarkable about the conjecture of \cite{Bern:2005iz} is that also the finite part of the amplitude, properly defined, does
exponentiate. Since the conjecture of \cite{Bern:2005iz} is an all-loop one, it should also be valid in the strong coupling regime,
$\lambda=g^2 N>>>1$, and can thus be tested by using the AdS/CFT correspondence \cite{Maldacena:1997re}. However, the objects which naturally occur in a conformal field theory are gauge invariant operators and their correlation functions and not the scattering amplitudes.

In a remarkable paper, the authors of \cite{Alday:2007hr} gave a prescription of how to calculate scattering amplitudes in the context of AdS/CFT. In particular, in order to simplify the boundary conditions of the problem, they performed a T-duality in four of the directions of the AdS space. As a result, it became evident that the calculation of the scattering amplitude is equivalent to the vacuum expectation value of polygonal Wilson loop whose contour is comprised of light-like segments, each segment corresponding to the momentum of a scattered gluon. Using this approach, they succeeded in finding the minimal surface for the case of a loop with four segments which corresponds to the logarithm  of the four gluon amplitude at strong coupling.
The expression they obtained is in agreement with the conjecture put forward in \cite{Bern:2005iz} , when one takes into account the value of the cusp anomalous dimension at strong coupling.

The aforementioned equivalence of the amplitude and the expectation value of the Wilson loop, although justified only at strong coupling, suggests that the same duality may hold at weak coupling too, order by order in perturbation theory.
That this is true at the one loop level was confirmed in \cite{Drummond:2007aua} for the four gluon amplitude and in \cite{Brandhuber:2007yx} for the n-gluon amplitude. Subsequently, the authors of \cite{Drummond:2007cf} were able to confirm this conjecture by computing the expectation value of the Wilson loops with four and five legs at two loops.

In the meanwhile, some doubt was cast on the validity of the BDS conjecture beyond one loop and for amplitudes with six gluons or more \cite{Alday:2007he}, \cite{Bartels:2008ce}. On the Wilson loop side, an important development was the derivation of the anomalous conformal Ward identities which the Wilson loop obeys \cite{Drummond:2007au}. By exploiting these identities it became apparent that the BDS conjecture is correct
and agrees with the Wilson  loop picture for four and five particles. However, for amplitudes with more than five gluons, one can construct conformally invariant cross ratios. As a consequence, one can add any function of these cross ratios to the BDS ansatz and still have the conformal Ward identities satisfied. This function which encodes the deviation from the BDS conjecture is termed as the finite remainder $R$. The authors of \cite{Drummond:2007bm} calculated the remainder of the two loop six-edged Wilson loop numerically, and found that it is different from zero. In a parallel development, the corresponding two loop six gluon MHV amplitude was calculated and it was directly verified that the BDS ansatz has to be modified \cite{Bern:2008ap}. The comparison of these results \cite{Drummond:2008aq}, \cite{Bern:2008ap}
 shows that the parity even finite part of the MHV amplitude and the Wilson loop are in agreement (up to a constant) for six particles at two loops.

As was, briefly, discussed above the Wilson loops obey conformal Ward identities which constrain to a great extent
their expectation values. The conformal symmetry related to these Ward identities is not the conformal symmetry of the
original space where the gluons live and scatter. It is a symmetry of a dual space where the Wilson loops live, and at strong coupling this space is the boundary of the AdS space obtained after the T-dualities are performed.
The nature of this symmetry, as well as its connection to integrability were clarified in \cite{Berkovits:2008ic},
\cite{Beisert:2008iq}.
The Wilson loop/amplitude duality suggests that the tree level scattering matrix of the ${\cal N}=4$ theory
could possess a dual supersymmetrised version of the dual conformal theory.
This was proved and further studied in \cite{Drummond:2008vq}, \cite{Drummond:2008bq}, \cite{Drummond:2008cr}, \cite{Brandhuber:2008pf}, \cite{ArkaniHamed:2008gz}.

In this paper we study the relation between null Wilson loops with a
self-crossing and their corresponding amplitudes\footnote{Other
  studies of the amplitude in particular kinematic limits (multi-Regge
  kinematics) include \cite{Bartels:2008sc},
  \cite{Brower:2008ia,Brower:2008nm}, \cite{DelDuca:2008jg}.}. One
important aspect of our analysis is that the kinematics of
self-intersecting Wilson loops are in the Minkowskian region.
Firstly, we compute the one loop contribution to the expectation
value of a self-intersecting loop. This computation shows that the
Wilson loop develops an additional $1/\epsilon$ pole associated to the
intersection. The Wilson loop/amplitude conjecture indicates that the
same $1/\epsilon$ singularity should appear at the corresponding
amplitude. We verify this by examining the finite part of the one-loop
amplitude at the corresponding kinematic limit.  Furthermore, we focus
on the behaviour of the finite remainder function $R$
\footnote{Recently, the strong coupling behaviour of $R$ for the case
  of regular polygons was studied in \cite{Alday:2009yn}.}.  By
exploiting the renormalisation group equations which govern the
dependence of the Wilson loops on the renormalisation scale $\mu$ we
argue that the remainder function $R(u_1,u_2,u_3)$ explodes in the
imaginary direction when approaching the self-intersecting Wilson
loop. This is similar to what happens at one loop, where the finite
part of the amplitude also diverges, and suggests that $R$ has a
branch cut in the negative $u_2$ axis when the other two ratios are
approximately equal $u_1\approx u_3$.

The rest of the paper is organised as follows. In Section 2, we
discuss the kinematics of a Wilson loop with a self crossing. In
Section 3, we compute the one loop contribution of the crossing to the
expectation value of a self-intersecting Wilson loop and compare with
the the corresponding one loop amplitude. In Section 4, we derive the
renormalisation group equations relevant for the self-intersecting
null Wilson loop and its implications for the remainder function $R$.
Finally, in Section 5 we comment on the results of the two previous
sections.

\sect{Kinematics of the self crossing Wilson loop}\label{kinematics}

As a first step, we derive some useful relations for the kinematic
configuration of a self-intersecting Wilson loop (see Figure 1). By
translational invariance one can choose the intersection point to be
at zero. We also denote the momenta of the intersecting gluons by
$p=p_4$ and $q=p_1$. $x$ and $y$ are the fractions of $p$ and $q$ from
zero to $x_5$ and $x_1$ respectively. $P$ is the sum of the momenta
from the tip of $q$ to the beginning of $p$ while $Q$ is the
corresponding sum from the tip of $p$ to the beginning of $q$.
\begin{eqnarray}\label{P^2}
P^2=x_{15}^2 \qquad Q^2=x_{24}^2
\end{eqnarray}
 Momentum conservation for the upper half of the loop gives:
\begin{eqnarray}\label{mc1}
xp+yq+P=0.
\end{eqnarray}
Similarly, for the lower half of the loop it gives:
\begin{eqnarray}\label{mc2}
(1-x)p+(1-y)q+Q=0.
\end{eqnarray}
By dotting \eqref{mc1} with  $p$ we get:
\begin{eqnarray}\label{4}
y=-\frac{P\cdot p}{ p\cdot q},
\end{eqnarray}
while by dotting \eqref{mc1} with $q$ we get:
\begin{eqnarray}\label{5}
x=-\frac{P\cdot q}{ p \cdot q}.
\end{eqnarray}
Similarly from \eqref{mc2} it is easy to obtain:
\begin{eqnarray}\label{6}
1-y=-\frac{Q\cdot p}{ p \cdot q}
\end{eqnarray}
and
\begin{eqnarray}\label{7}
1-x=-\frac{Q\cdot q}{ p \cdot q}.
\end{eqnarray}
Furthermore, \eqref{mc1} and \eqref{mc2}
can be solved with respect to $P$ and $Q$
and squared to give:
\begin{eqnarray}\label{8}
P^2=2xy p\cdot q,
\end{eqnarray}
\begin{eqnarray}\label{9}
Q^2=2(1-x)(1-y) p\cdot q.
\end{eqnarray}
Finally, with the help of \eqref{8},\eqref{4} and \eqref{9},\eqref{6} we get:
\begin{eqnarray}\label{10}
s\doteq (P+p)^2=-2y(1-x)p\cdot q
\end{eqnarray}
and
\begin{eqnarray}\label{11}
t\doteq (P+q)^2=(Q+p)^2=-2x(1-y)p\cdot q
\end{eqnarray} respectively.
By using the equations listed above it is straightforward to verify
that:
\begin{eqnarray}\label{12}
P^2 Q^2=st=4xy(1-x)(1-y )p\cdot q.
\end{eqnarray}
This is one of the key relations of this Section.
In what follows we will also need the relation:
\begin{eqnarray}\label{13}
P^2+ Q^2-s-t=2 p\cdot q.
\end{eqnarray}

Before going on, let us elaborate  on the kinematics of the loop
appearing in Figure 1.
The first thing one can observe is that the value of the loop depends on
six independent Lorentz invariant variables. One way to see this is as follows:
The two momentum conservation relations $(1-x)p+(1-y)q+p_2+p_3=0$
and $xp+yq+p_5+p_6=0$ allow us to express two of the momenta, say
$p_5$ and $p_3$ in terms of $p,q,p_2,p_6$ and $x,y$.
Thus, we are left with four
independent momenta plus the pair $(x,y)$.
From the four independent momenta one can build six Lorentz invariant
combinations $p_i \cdot p_j, i,j=1,2,4,6$. However, one should remember
to impose the conditions that $p_5$ and $p_3$ are massless. These conditions
$p_5^2=0$ and $p_3^2=0$ supply two relations between the six  Lorentz invariant
combinations mentioned above. As a consequence, one has 4 Lorentz invariants
plus $x,y$ which makes six invariants in total.

In what follows, we express the nine variables appearing
in a six-edged Wilson loop in terms of six quantities:
\begin{eqnarray}\label{invariants}
x_{15}^2=\frac{xy}{(1-x)(1-y)} s_2 \qquad x_{24}^2= s_2 \qquad
x_{14}^2=-\frac{ys_2}{1-y} \qquad x_{25}^2=-\frac{xs_2}{1-x}\nonumber\\
x_{13}^2=\frac{t_2}{1-y} \qquad x_{26}^2=\frac{t_1}{y}\qquad
x_{35}^2=\frac{t_2}{1-x} \qquad x_{46}^2=\frac{t_1}{x}\nonumber \\
x_{36}^2=\frac{t_2}{1-y}+\frac{t_1}{y}+2p_2 \cdot p_6,
\end{eqnarray}
 where
\begin{eqnarray}\label{t1t2}
 s_1\doteq 2xy p\cdot q \qquad s_2\doteq 2(1-x)(1-y) p\cdot q \nonumber \\
 t_1\doteq 2 y q \cdot p_6 \qquad t_2\doteq 2(1-y)q \cdot p_2.
\end{eqnarray}
Let us mention that $s_1,t_1$ and $s_2,t_2$ are the invariants of the
loops $C_1$ and $C_2$ respectively (see Figure 1b).
Using the relations in \eqref{invariants} one can deduce the
values of the three cross ratios in terms of six independent quantities
$t_1,t_2,s_2,p_2 \cdot p_6,x,y$. These read:
\begin{eqnarray}\label{crossratios}
u_1=\frac{x_{13}^2x_{46}^2 }{x_{14}^2x_{36}^2}=
\frac{-t_1t_2}{xy s_2(\frac{t_2}{1-y}+\frac{t_1}{y}+2p_2 \cdot p_6)} \nonumber \\
u_3=\frac{x_{35}^2x_{26}^2 }{x_{36}^2x_{25}^2}=u_1\qquad
u_2=\frac{x_{24}^2x_{15}^2 }{x_{14}^2x_{25}^2}=1.
\end{eqnarray}
Thus, we see that for the loop shown in Figure 1 there is only one free cross ratio,
 since 2 of them are equal $u_1=u_3$ and the third is equal to one, $u_2=1$.
 \begin{figure}[t]
   \centering
  \includegraphics[width=0.6\textwidth]{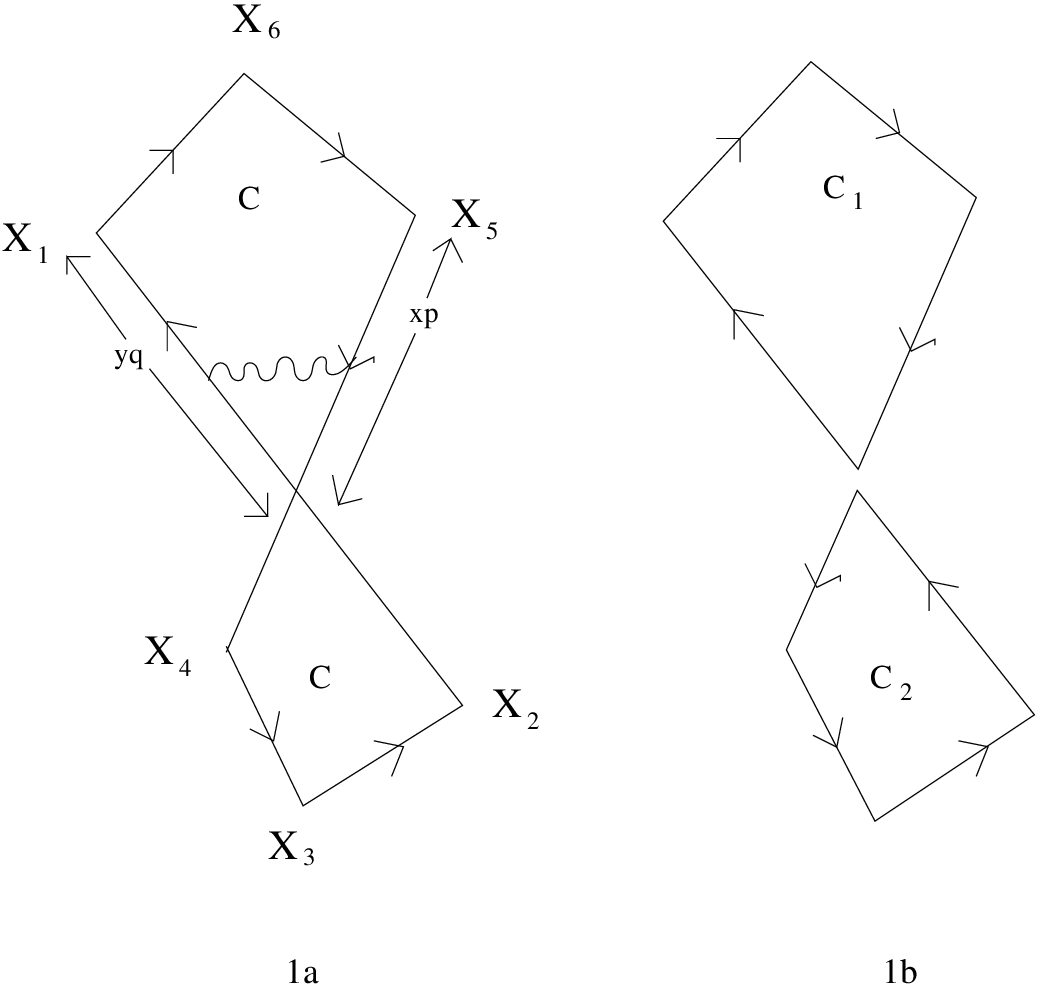}
   \caption{Wilson loop configurations which mix under renormalisation. Diagram 1a depicts the contour $C$ of the Wilson loop $W_1=\frac{1}{N}{\rm Tr} \, \mathcal{P} \exp \left[ i g\oint_{C} \! d\tau \Big(  A_\mu (x(\tau )) \dot{x}^{\mu} (\tau )
 \Big) \right]$ while diagram 1b depicts the contour $C_1\cup C_2$ of the
Wilson loop $W_2=W(C_1)W(C_2)$. The relation between the external gluon momenta and the $x$-coordinates
  is $p_i=x_i-x_{i+1}$. The wiggly lines denote gluon propagators. Diagram 1a gives a contribution to the one loop value of the Wilson
 loop with a self crossing.}
 \end{figure}

\sect{One loop result for a Wilson loop with a self-crossing and the BDS conjecture}
In this section, we calculate the one loop corrections to the vacuum
expectation value of the Wilson loop depicted in Figure 1a. In ${\cal N}=4$ SYM
the form of this operator is given by \cite{Drukker:1999zq}:
\beq
\label{wil}
W(C)  \ \doteq \frac{1}{N}\ {\rm Tr} \, \mathcal{P} \exp \left[ i g\oint_{C} \! d\tau \Big(  A_\mu (x(\tau )) \dot{x}^{\mu} (\tau ) +
\phi_i  (x (\tau)) \dot{y}^i (\tau) \Big) \right]
\ ,
\eeq
where $\mathcal{P}$ is the usual path-order symbol
and $x^{\mu}(\tau)$ and $y^i(\tau)$ parametrise the loop $C$ \footnote{Although a number of results have been
already obtained for other observables in deformed theories \cite{deform} or theories in less dimensions, e.g. ABJM,
  there are no results for  Wilson operators with a self-crossing in these theories. It would be interesting to perform
  such calculations along the lines of the following Sections. Furthermore, it would be interesting to clarify the connection
  between the expectation values of Wilson operators and the infinite spin limit of n-point correlators in the $SL(2)$ sector of $N=4$ SYM \cite{corr}.}.
It is possible, if one wishes, to include the fermions and use
additional functions of $\tau$ needed to parametrise the trajectory of
a particle in superspace. In what follows, we  focus on bosonic loops
by setting all these functions to zero. In addition we set $\dot{y}^i=0$.
As a result, the Wilson loop is  locally supersymmetric $\dot{x}^2=\dot{y}^2$
only if  $\dot{x}^2=0$. For the configurations of Figure 1 this is true since
the segments building the loop are null.

Effectively, what we are left with is the form of the Wilson operator in QCD.
The properties of these operators are well-studied \cite{wil1,wil2,wil3}.
It is known that the expectation value of a smooth loop without any
intersection is finite when expressed in terms of the renormalised
parameters \cite{wil1,wil2}. If the loop is not smooth, but has one or more cusps,
then it is no longer finite since additional ultraviolet (UV)
divergences make their appearance. However, it can be shown
that the divergences can be renormalised multiplicatively
(as long as there are no cusps that lie on the light cone) \cite{wil1,wil2}.

The situation is more intricate when the loop intersects itself.
In this case, the loop functions belonging to the set of loops which are the same
as the original loop except at the crossing points mix with each other under
renormalisation promoting the renormalisation constant $Z$ to a matrix \cite{wil3}.
For example, the two configurations appearing in Figure 1 mix
under renormalisation. Let us note, that intersecting Wilson loops are important since
it is precicely these configurations which give the quantum corrections to the
Migdal-Makeenko loop equations \cite{Migdal}.

As soon as one considers self-intersecting loops which are built
from null segments an apparent puzzle arises. As we will show in the rest of this
Section, the Wilson loop of Figure 1 has an $1/\epsilon$ singularity
associated to the crossing \footnote{The results of this section are not in agreement with those of
 Section 6.1 of \cite{McGreevy:2007kt}.}. The direct connection
 between the Wilson loop and the corresponding gluon amplitude,
 discussed in the Introduction, implies that the amplitude should also
have the same $1/\epsilon$ pole.
But the IR singularities of the amplitudes are well-known.
At large $N$ they come from the exponentiation of the soft and
collinear singularities and due to planarity the involve only two particle
invariants of adjacent gluons \cite{Bern:2005iz}. Since, as we will shortly see,
the Wilson loop UV singularity depends on $s_{pq}$ it seems that there is
a disagreement between the two quantities. We will discuss the
resolution of this puzzle at the end of this Section.

Next, we proceed to the one loop correction to the Wilson loop of Figure 1a.
As discussed in \cite{Brandhuber:2007yx} there are two types of
non-vanishing diagrams. In the first one,
a gluon joins two adjacent segments meeting at a cusp. This kind of diagrams
give the anticipated $1/\epsilon^2$ pole when both ends of the gluon approach
the cusp.

The second type consists of those diagrams where the gluon  stretches
 between two non-adjacent segments.
Generically, this class of diagrams gives a finite
contribution even when evaluated in four dimensions.
Our calculation is similar to the one of \cite{Brandhuber:2007yx}
except that now one has to be careful when the gluon is
exchanged between the lines carrying momentum $p$ and $q$.
In this case one can split the integral in four pieces accounting for the
 four different ways in which the gluon can be exchanged.

To start, we parametrise the crossing lines as
\beqa\label{param}
 x(\tau_p)=x_5+\tau_p p \\
 x(\tau_q)=x_2+\tau_q q,
\eeqa
where $0 \le \tau_p,\tau_q \le 1$.
\eqref{param} implies that
\beqa\label{x-x}
 x(\tau_p)-x(\tau_q)=x_5-x_2+\tau_p p-\tau_q q=P+q+\tau_p p-\tau_q q.
\eeqa
One can now define the new variables $b=-(1-y)+\tau_q$ and $a=x- \tau_p$ and
use  momentum conservation  \eqref{mc1} to rewrite \eqref{x-x} as
\beqa\label{x-xf}
 x(\tau_p)-x(\tau_q)=-b q-a p.
\eeqa
It is now straightforward to evaluate the integral
 in $D=4-2 \epsilon$ dimensions.
An important comment is in order. The propagator we are using is:
\beqa\label{prop}
\Delta_{\mu \nu}(x)=-\frac{\pi^{\epsilon}}{4 \pi^2} \Gamma(1-\epsilon)
\frac{\eta_{\mu \nu}}{(-x^2-i0)^{1-\epsilon}}.
\eeqa
Notice that the prescription for avoiding the poles of
the Feynman propagator is opposite
to that in configuration space and opposite to that of
\cite{Drummond:2007aua,Brandhuber:2007yx}.
This is because only then the analytic
properties of the divergent part of the amplitude
as obtained from the 2 mass easy box integral are the same as the analytic
properties of
the expression obtained from a cusp in the Wilson loop calculation. Namely,
both should behave as:
\beqa
-\frac{1}{\epsilon^2}\Big(\frac{-s-i0}{\mu^2}\Big)^{\epsilon}.
\eeqa
Somehow, the prescription in \eqref{prop} is natural to impose,
since the Wilson loop actually lives in momentum space.

The finite part of the Wilson loop expectation value originating from
a gluon exchange between the crossing momenta $p$ and $q$ is:

\beqa\label{int}
F_{\epsilon}=\frac{1}{N}\frac{-(i g\mu^{\epsilon})^2\Gamma(1-\epsilon)}{4 \pi^{2-\epsilon}}
 \int_0^1 \, d\tau_p\int_0^1 \, d\tau_q
  \frac{p \cdot q}{(-(x(\tau_p)-x(\tau_q))^2-i0)^{1-\epsilon}}\frac{N^2}{2}
\nonumber\\=
  \frac{N}{2}\frac{-(i g\mu^{\epsilon})^2\Gamma(1-\epsilon)}{4 \pi^{2-\epsilon}}
   \int_{-(1-x)}^x \, da \int_{-(1-y)}^y \, db
    \frac{p \cdot q}{(-2 ab \,p \cdot q -i0)^{1-\epsilon}}.
\eeqa
Notice that we have kept explicitly the prescription for the
Feynman propagator.
Then the integral becomes:
\beqa\label{int2}
F_{\epsilon}=&&\frac{N}{2}\frac{-( g\mu^{\epsilon})^2 \Gamma(1-\epsilon)}{4 \pi^{2-\epsilon}}
\frac{1}{2\epsilon^2}
\Big[(-s_{pq}xy-i0)^{\epsilon}
+(-s_{pq}(1-x)(1-y)-i0)^{\epsilon}\nonumber\\ &&-(s_{pq}x(1-y)-i0)^{\epsilon}-(s_{pq}y(1-x)-i0)^{\epsilon}
\Big]
\eeqa
Without loss of generality one can assume that $p \cdot q>0$.
The expansion in $\epsilon$ of the square bracket in \eqref{int2} gives
\beqa\label{square}
\epsilon(-2\pi i)+\frac{1}{2}\epsilon^2\big(\log^2(-s_{pq}xy-i0)+
\log^2(-s_{pq}(1-x)(1-y)-i0)\nonumber\\
-\log^2(s_{pq}x(1-y)-i0)-\log^2(s_{pq}y(1-x)-i0)\big)
+\mathcal{O}(\epsilon^3).
\eeqa
In all the manipulations above we have put the  branch cut of the logarithm
on the negative real axis. Using \eqref{square} our result becomes
\beqa\label{intfin}
F_{\epsilon}&=&\frac{g^2 N}{2}\frac{\Gamma(1-\epsilon)}{4 \pi^{2-\epsilon}}
 \frac{\mu^{2\epsilon}}{\epsilon}( \pi i)
-\frac{g^2 N}{8}\frac{1}{4 \pi^{2}}\big(\log^2(-s_{pq}xy-i0)+
\log^2(- s_{pq}(1-x)(1-y)-i0)\nonumber\\
&&-\log^2(s_{pq}x(1-y)-i0)-\log^2(s_{pq}y(1-x)-i0)\big)
+\mathcal{O}(\epsilon).
\eeqa

We now turn to the puzzle mentioned earlier in this Section.
It is well-known that the complete one loop amplitude is expressed
as a sum of two-mass easy box functions, all having coefficient equal to one
 \cite{Bern:1994zx}, \cite{Brandhuber:2004yw}. Although each box function contains poles in
$\epsilon$ whose coefficient depends on multi-particle invariants,
it happens that after performing the sum of all the two-mass easy box functions the infrared divergent terms involve
only two particle invariants of adjacent particles. However, our expression
for the Wilson loop contains a pole that depends on the invariant
of two non-adjacent gluons, apparently not present in the amplitude.
The resolution of this puzzle is the following.
The finite part of the two mass easy box whose massless legs are $p$ and $q$
has a $1/\epsilon$ pole identical to this of \eqref{int2}.
This can be seen by writing the all orders in $\epsilon$ expression for the
finite part, which can be found in \cite{Brandhuber:2005kd}. This reads:
\beqa\label{allorders}
&&F^{\rm 2me} (s, t, P^2, Q^2) =
-\frac{1}{2\epsilon_{IR}^2}
\left[
 \Big( \frac{c \mu_{IR}^{2}}{1-c(P^2+i0)} \Big)^{\epsilon_{IR}} \,
\mbox{}_{2}F_1 \left( \epsilon_{IR}, \epsilon_{IR}, 1+ \epsilon_{IR}, \frac{1}{ 1 - c (P^2+i0) }\right)
\right.\nonumber\\
&&\left.
  +
\Big( \frac{c \mu_{IR}^{2}}{1-c(Q^2+i0)} \Big)^{\epsilon_{IR}} \,
\mbox{}_{2}F_1 \left( \epsilon_{IR}, \epsilon_{IR}, 1+ \epsilon_{IR}, \frac{1}{1 - c (Q^2+i0) }\right)
 - \,
\right.\\
&&\left.
\Big( \frac{c \mu_{IR}^{2}}{1-c(s+i0)} \Big)^{\epsilon_{IR}} \,
\mbox{}_{2}F_1 \left( \epsilon_{IR}, \epsilon_{IR}, 1+ \epsilon_{IR}, \frac{1}{1 - c (s+i0) }\right)\, - \,
\right.\nonumber\\
&&\left.
 \Big( \frac{c \mu_{IR}^{2}}{1-c(t+i0)} \Big)^{\epsilon_{IR}} \,
\mbox{}_{2}F_1 \left( \epsilon_{IR}, \epsilon_{IR}, 1+ \epsilon_{IR}, \frac{1}{1 - c (t+i0) }\right)\right]
\ ,
\nonumber
\eeqa
where
\beqa\label{c}
c=\frac{P^2+Q^2-s-t}{P^2 Q^2-st}.
\eeqa
By looking at \eqref{12}, \eqref{13} it is immediate to see that
$c\rightarrow - \infty$ as one approaches the
crossing configuration \footnote{One can show that $c\leq 0$.}.
By taking into account that
 $\epsilon=-\epsilon_{IR}>0$ and $\mu_{IR}= \mu^{-1}$
one can multiply \eqref{allorders} by $a=\frac{g^2N}{8 \pi^2}$
and take its limit as $c\rightarrow - \infty$
to obtain \eqref{int2} \footnote{\label{foot}The $\mu$ appearing
in the Wilson loop calculation can be redefined as $\mu^2 \pi e^{\gamma_E}\rightarrow \mu^2$ (see \cite{Drummond:2007aua}) to absorb part of the Gamma function and a factor of $\pi^{\epsilon}$ which are present in \eqref{int2}. After this redefinition the two expressions agree up to order $\mathcal{O}(\epsilon^0)$.
Furthermore, $\mu_{IR}$ appearing in \eqref{allorders} is related to the dimensional regularisation scale $\mu_{amp}$  used in the calculation of the amplitude by $\mu_{IR}^2=4 \pi e^{-\gamma}\mu_{amp}^2.$}, since in this limit the hypergeometric functions
become one. Furthermore, it is not difficult to verify that all the other
finite contributions originating from the gluon exchange
between two non-adjacent segments do remain finite.

It is instructive to consider the behaviour of the finite part
by first setting $\epsilon$ to zero and then approach the
crossing configuration. In this case the finite part reads
\cite{Duplancic:2000sk,Brandhuber:2007yx}:
\beqa\label{e=0}
F_{\epsilon=0}=\frac{1}{2}\big(-Li_2(1-c (s+i0))-Li_2(1-c (t+i0))+Li_2(1-c (P^2+i0))\nonumber \\
+Li_2(1-c (Q^2+i0))\big).
\eeqa
In the limit $c\rightarrow - \infty$
one can use the identity
\beqa\label{ID}
Li_2(1-x)+Li_2(1-x^{-1})+\frac{1}{2}\log^2(x)=0
\eeqa
to obtain
\beqa\label{app}
Li_2(1-c (s+i0))=-\frac{1}{2}\log^2(c(s+i0))-\frac{\pi^2}{6}.
\eeqa
Employing \eqref{app} we can rewrite \eqref{e=0} as
\beqa\label{e=0app}
F_{\epsilon=0}=\frac{1}{4}\big(\log^2(c(s+i0))+\log^2(c(t+i0))
-\log^2(c(P^2+i0))-\log^2(c(Q^2+i0))\big).\nonumber\\
\eeqa
Given that we have chosen the regime where $P^2,Q^2>0$ while $s,t<0$
we can rewrite \eqref{e=0app} as
\beqa\label{e=0appfin}
&&F_{\epsilon=0}=\frac{1}{4} \Big( (\log(-c)+\log(-s-i0))^2 +(\log(-c)+\log(-t-i0))^2\nonumber\\
&&-(\log(-c)+\log(-P^2-i0))^2
-(\log(-c)+\log(-Q^2-i0))^2 \Big)\nonumber\\
&&=\log(-c)i\pi+\frac{1}{4}\Big(\log^2(-s-i0)
+\log^2(-t-i0)
-\log^2(-P^2-i0)-\log^2(-Q^2-i0)\Big).\nonumber\\
\eeqa
It is amusing to notice that \eqref{e=0appfin}
is what one has in \eqref{intfin} \footnote{After multiplying \eqref{e=0appfin}  by $a=\frac{g^2 N}{8 \pi^2}$. See also footnote \ref{foot}.}
 after the identification
$log(-c)\leftrightarrow \mu^{2\epsilon}/\epsilon$ which implies
$-log(1-u_2)\leftrightarrow 1/\epsilon$.

In fact, it is natural to expect this relation
since one is dealing with logarithmic singularities.
One can view the situation where
$c\rightarrow -\infty$ as another way to regularise the crossing singularity.
This can be seen by considering the simple case where the crossing singularity is regularised
by separating the two gluon momenta $p$ and $q$ by a
space-like distance $z^{\mu}\rightarrow 0$ satisfying
$p \cdot z=q \cdot z=0$ and  $z^2=-\vec{z}^2$. In this case,
one can parametrise
the momenta by $x_p^{\mu}=a p^{\mu},x_q^{\mu}=b q^{\mu}+ z^{\mu}$
where $a,b\in[-1/2,1/2]$.
It is then straightforward to evaluate $P^2=Q^2=p \cdot q/2-\vec{z}^2$ and
$s=t=-p \cdot q/2-\vec{z}^2$ from which one deduces that
$c=-\frac{1}{\vec{z}^2}$. It is easy to calculate
the gluon exchange integral of the aforementioned configuration whose
$z^{\mu}\rightarrow 0$ limit has the leading behaviour
$- \pi i \log(\vec{z}^2)= \pi i \log(-c)$ in agreement with
\eqref{e=0appfin}.

It will be useful, in what follows, to notice that one can split $-c$ in two
pieces, one depending on the conformal ratio $u_2$ only and another
which is not a function of the cross ratios. Namely,
\beqa\label{cu}
\log(-c)=\big(\log(\frac{P^2+Q^2-s-t}{st})-\log(1-u_2)\big).
\eeqa

In the next section we derive the renormalisation group equations obeyed by
null cusped Wilson loops with a self-crossing and discuss their implications for the
two loop remainder function $R(u_1,u_2,u_3)$.

\sect{Renormalisation group equations and the finite remainder $R$}
As was briefly mentioned in the previous section,
the renormalisation properties of self intersecting Wilson loops are governed
by the cross anomalous dimension matrix which in our case will be
a gauge invariant $2\times 2$ matrix depending only on the coupling constant
and crossing angle $\gamma$.
In particular, the dependence of the renormalised expressions for the loops of Figure 1 \footnote{To be precise the RGE written below govern the evolution of self-intersecting loops where neither the cusps nor the self-crossing have edges that lie on the light-cone. The RGE relevant for Figure 1 will be derived in the rest of this Section.}, $W_1^r$ and $W_2^r$,
on the renormalisation point $\mu$ is given by the renormalisation group (RG)
equations \cite{wil3}, \cite{Korchemskaya:1994qp}
\beqa\label{RG}
(\mu \frac{\partial}{\partial \mu}+\beta(g)\frac{\partial}{\partial g})W_i^r
=-\Gamma^{ij}(\gamma,g)W_j^r-\sum_k \Gamma_{cusp}(\gamma_k,g)\delta_{ij}W_j^r ,\ i,j=1,2,
\eeqa
where
\beqa\label{beta}
\beta(g)=\lim_{\epsilon\rightarrow 0}\frac{\partial g(g_B,\mu,\epsilon)}{\partial \mu},
\eeqa
$g_B$ is the bare coupling constant which is kept fixed as we differentiate
the renormalised coupling
$g$ with respect to $\mu$ and the sum over $k$ is a sum
over all the cusps of figure 1a.

\eqref{RG} holds for generic loop configuration and implies that the Wilson loops
are multiplicatively renormalised. However, one should be careful when the tangent
 of the lines at the crossing point and/or at one or more of the cusps lie on the light-cone. This is precisely
the case for the loops of Figure 1. As was pointed out in \cite{trick}, for the Wilson loops that contain both
light-cone and cusp divergences there is no multiplicative renormalisability and, as a result,
\eqref{RG} does not hold any more.
Before going on, let us elaborate on the form  the RG equations take in such a case.

The cross anomalous dimension matrix for a generic crossing configuration in QCD
is known up to two loop order and is given in \cite{Korchemskaya:1994qp}.
Although what interest us is the expectation value of the Wilson loops at
large $N$ we remain general and keep, for the moment, the number of the colours $N$ finite.
Since we eventually want to focus on the loops of Figure 1, where $p$ and $q$ lie
on the light-cone, it is sufficient to consider the large $\gamma$ behaviour of the
crossing matrix. The general structure of this matrix at large $\gamma$ is that of equation 5.48 of
\cite{Korchemskaya:1994qp}. $\Gamma_1$ is a function which accepts
a perturbative expansion in the coupling constant $\Gamma_1=\Gamma_1^{(1)}a+\Gamma_1^{(2)}a^2+...$.
The one-loop coefficient $\Gamma_1^{(1)}$ is determined by the exchange of a single gluon and
it is, thus, the same in both QCD and $N=4$ SYM. Its value, $\Gamma_1^{(1)}=2$,
can be read directly from \cite{Korchemskaya:1994qp}. However, this is not the case for
$\Gamma_1^{(2)}$ since the determination of its value requires
the self-energy correction to the gluon which is different in different theories.

However, one word of caution is in order. Firstly, equation 5.48 of \cite{Korchemskaya:1994qp}
was obtained by using the $-x^2+i0$ prescription for the Feynman propagator.
We have seen in Section 3 that in order to have agreement between
the analytic properties of the expression for the cusp and the
corresponding expression for the IR divergences of the amplitude, one has to
 adopt the opposite prescription, namely $-x^2-i0$.
The effect of this change on the cross anomalous dimension matrix is to substitute
$i \pi$ with $-i \pi$ wherever it appears in the matrix.
A second observation concerns the normalisation of the loop functions.
In our conventions the loop $W_1$ is normalised by multiplying it by $1/N$
see \eqref{wil}, while $W_2$ is normalised by multiplying it by $1/N^2$.
On the other hand, these normalisations are absent from the Wilson line definitions of \cite{Korchemskaya:1994qp}.
As a result, the values for the entries of the crossing matrix we have to use
are modified with respect to those of \cite{Korchemskaya:1994qp}
$\Gamma_{11}^{(Kor.)}$, the modification being
$\Gamma_{11}^{ours}=\Gamma_{11}^{(Kor.)},\Gamma_{12}^{ours}=N \Gamma_{12}^{(Kor.)}$,
$\Gamma_{21}^{ours}=\frac{1}{N}\Gamma_{21}^{(Kor.)},\Gamma_{22}^{ours}=\Gamma_{22}^{(Kor.)}$.

With these observations taken into account the cross anomalous dimension matrix reads:
  \begin{gather}
    \Gamma =
    \begin{pmatrix}
      \frac{i \pi}{N^2} \Gamma_1(g) & -i \pi \Gamma_1(g)\\
      -\frac{\gamma\Gamma_{cusp}(g)}{N^2}-\frac{i \pi}{N^2}(2\Gamma_{cusp}(g)-\Gamma_1(g)) &\gamma\Gamma_{cusp}(g)+\frac{i \pi}{N^2}(2\Gamma_{cusp}(g)-\Gamma_1(g))
      \end{pmatrix}.
\end{gather}

By keeping only the leading terms in $\gamma$ in each of the entries of the crossing matrix it is possible to rewrite
\eqref{RG} as
\beqa\label{RG1}
(\mu \frac{\partial}{\partial \mu}+\beta(g)\frac{\partial}{\partial g})W_1^r
=-\frac{i \pi}{N^2}\Gamma_1(g)(W_1^r-N^2W_2^r)-\sum_k \Gamma_{cusp}(\gamma_k,g)W_1^r&&\nonumber\\
(\mu \frac{\partial}{\partial \mu}+\beta(g)\frac{\partial}{\partial g})(W_1^r-N^2W_2^r)
=\Big(-\frac{i \pi}{N^2}\Gamma_1(g)-\gamma
\Gamma_{cusp}(g)&&\nonumber\\
-\sum_k \Gamma_{cusp}(\gamma_k,g)\Big)
(W_1^r-N^2W_2^r).&&
\eeqa
When $p$ and $q$ lie on the light-cone $\gamma$ becomes infinite and \eqref{RG1} meaningless.
In order to find the equations the loop functions satisfy in this case, one can use a similar trick
to the one used in \cite{trick} for a loop containing a null cusp.
One can take the second equation of \eqref{RG1} and after dividing both sides by $W_1^r-N^2W_2^r$
 differentiate it with respect to $-2 p \cdot q$ to get \footnote{In what follows,
 we also use the fact that the large $\gamma_k$
 behaviour of $\Gamma_{cusp}(\gamma_k,g)$ is given by $\Gamma_{cusp}(\gamma_k,g)=\frac{\gamma_k}{2} \Gamma_{cusp}(g)$,
 where $\gamma_k=\log\frac{2p_k \cdot p_{k+1}}{\sqrt{p_k^2}\sqrt{p_{k+1}^2}}$ \cite{Rad}.}
\beqa\label{RG2}
\frac{\partial}{\partial (-2 p \cdot q)}
(\mu \frac{\partial}{\partial \mu}+\beta(g)\frac{\partial}{\partial g})\log(W_1^r-N^2W_2^r)
=-\Gamma_{cusp}(g) \frac{1}{-2 p \cdot q-i0}.
\eeqa
Notice that the right hand side of \eqref{RG2} is well defined even when
$p^2=q^2=0$. Strictly speaking  \eqref{RG2} holds for $2 p \cdot q>>p^2q^2$.
We can now set $p^2=q^2=0$ on both sides of \eqref{RG2} and integrate back to get
\beqa\label{RG3p}
(\mu \frac{\partial}{\partial \mu}+\beta(g)\frac{\partial}{\partial g})\log(W_1^r-N^2W_2^r)
=-\Gamma_{cusp}(g) \log(\mu^2(-2 p \cdot q-i0))-\bar\Gamma(g,\gamma_k),
\eeqa
where $\bar\Gamma(g,\gamma_k)$ is an integration ''constant'' that depends on
the angles of the cusps but not on $s_{pq}$. One can then, repeatedly, differentiate with respect
 to the invariants associated to the cusps of Figure 1b to obtain
 \beqa\label{RG3}
 (\mu \frac{\partial}{\partial \mu}+\beta(g)\frac{\partial}{\partial g})\log(W_1^r-N^2W_2^r)
=-\Gamma_{cusp}(g) \log(\mu^2(-2 p \cdot q-i0))\nonumber\\
-\frac{1}{2}\sum_k \Gamma_{cusp}(g)\log(-\mu^2s_{k,k+1}-i0)
-\bar\Gamma(g),
 \eeqa
where $\bar\Gamma(g)$ is an integration constant to be determined from
plugging the explicit expressions of the loop functions in \eqref{RG3}.

One can apply the same trick on the first equation of \eqref{RG1}.
Namely we divide the first equation of \eqref{RG1} by $W_1^r$ and differentiate
repeatedly with respect to $s_{pq}$ and with respect to the invariants $s_{k',k'+1}$
associated to the cusps of Figure 1a.
The resulting equation together with \eqref{RG3} are the
proper renormalisation group equations describing the evolution of the loop
 functions involving a null self-crossing. Namely,
 \beqa\label{RGf}
(\mu \frac{\partial}{\partial \mu}+\beta(g)\frac{\partial}{\partial g})W_1^r
=-\frac{i \pi}{N^2}\Gamma_1(g)(W_1^r-N^2W_2^r)\nonumber\\
-\frac{1}{2}\sum_{k'}\Gamma_{cusp}(g)\log(\mu^2(-s_{k',k'+1}-i0))W_1^r-\tilde{\Gamma}(g)W_1^r \nonumber\\
(\mu \frac{\partial}{\partial \mu}+\beta(g)\frac{\partial}{\partial g})(W_1^r-N^2W_2^r)
=\Big(-\Gamma_{cusp}(g) \log(\mu^2(-2 p \cdot q-i0))-\bar\Gamma(g)\nonumber\\
-\frac{1}{2}\Gamma_{cusp}\sum_k \log(-\mu^2 s_{k,k+1}-i0)\Big)(W_1^r-N^2W_2^r).
\eeqa

The last set of equations are general and hold for any number of colours.
What really interests us is their behaviour as $N\rightarrow \infty$.
In such a limit  \eqref{RGf} becomes
 \begin{multline}
\label{RGinfty}
(\mu \frac{\partial}{\partial \mu}+\beta(g)\frac{\partial}{\partial g})W_1^r
=i \pi\Gamma_1(g)W_2^r-\Big(
\frac{1}{2}\sum_{k'}\Gamma_{cusp}(g)\log(\mu^2(-s_{k',k'+1}-i0))+\tilde{\Gamma}(g)\Big)W_1^r\\
(\mu \frac{\partial}{\partial \mu}+\beta(g)\frac{\partial}{\partial g})W_2^r
=\Big(-\Gamma_{cusp}(g) \log(\mu^2(-2 p \cdot q-i0))-\bar\Gamma(g)\Big)W_2^r\\
-\frac{1}{2}\sum_k\Gamma_{cusp}(g)\log(\mu^2(-s_{k,k+1}-i0)))W_2^r.
 \end{multline}
$\tilde{\Gamma}(g)$ and $\bar\Gamma(g)$ appear as constants of integration and they do not depend
on any of the kinematical invariants. This set of equations and in particular
the first one will allow us to draw conclusions about the behaviour of the remainder function $R$ near the crossing configuration of Figure 1a.

Before proceeding let us examine the second equation of \eqref{RGinfty}.
It is a well-known fact that at large $N$ the Wilson loop $W_2$ factorises into the product
of the expectation values of the two separate loops $C_1$ and $C_2$, that is
\beqa\label{fact}
W_2=\langle W(C_1)W(C_2)\rangle=\langle W(C_1)\rangle\langle W(C_2)\rangle.
\eeqa
But the result for $W(C_1)$ and $W(C_2)$ are known to all orders in perturbation theory
since they are four-sided null loops and their form is completely determined by dual conformal invariance.
 By using their expressions given in \cite{Drummond:2007cf} it is not difficult to verify that $W_2$
satisfies, indeed, the second renormalisation group equation of \eqref{RGinfty}.
It is important to note the lack of a factor of $1/2$
in front of the term containing $\log(\mu^2(-2 p \cdot q-i0))$ in the second equation of \eqref{RGinfty}.
This is the case because, as can be easily seen from  Figure 1b, there are two cusps having an  invariant proportional to $s_{pq}=2p \cdot q$.

The plan of the rest of this Section is to substitute the expressions for the renormalised Wilson loops $W^r_1$ and
$W^r_2$ in the right hand side of the first equation of\eqref{RGinfty} and solve for the dependence of $R$ on the scale $\mu$.
To this end, let us  recall the expressions for $W^r(C_1)$ and $W^r(C_2)$.
These are obtained by subtracting the poles
from the dimensionally regularised expressions for $W(C_1)$ and $W(C_2)$.
\beqa\label{W1W2}
W^r(C_1)=&&1-\frac{1}{4}\big(\Gamma_{cusp}^{(1)}a (\log^2(-\mu^2 t_1-i0)+\log^2(-\mu^2xy s_{pq}-i0))\big)+F_1+\mathcal{O}(\alpha^2)\nonumber\\
W^r(C_2)=&&1-\frac{1}{4}\big(\Gamma_{cusp}^{(1)}a (\log^2(-\mu^2 t_2-i0)+\log^2(-\mu^2(1-x)(1-y) s_{pq}-i0))\big)
\nonumber\\
+&&F_2+\mathcal{O}(\alpha^2),
\eeqa
where $t_1$ and $t_2$ are defined in \eqref{t1t2}.
$F_1$ and $F_2$ are the one-loop finite contributions to the upper and
lower half loops of Figure 1b, respectively. Their expressions can be found in \cite{Bern:2005iz}.

The expression for a generic six-edged loop, as obtained in dimensional
regularisation can be written as:

\beqa\label{W_1in}
\log W_1=\sum_{l=1}^2 \Big(a^l f_{WL}^{(l)}(\epsilon) w^{(1)}(l\epsilon)
+C_{WL}^{(l)}\Big)+R
\eeqa
In \eqref{W_1in} $w^{(1)}(l\epsilon)$ denotes the one loop contribution to the
Wilson loop evaluated at $4-2 l \epsilon$ dimensions while
$f_{WL}^{(l)}(\epsilon)=f_{0,WL}^{(l)}+f_{1,WL}^{(l)}\epsilon+f_{2,WL}^{(l)}\epsilon^2 $. The values for the constants $f_{0,WL}^{(2)}$,  $f_{1,WL}^{(2)}$ and
 $f_{2,WL}^{(2)}$ can be read from $f_{WL}^{(2)}(\epsilon)=-\zeta_2-7 \zeta_3\epsilon-5 \zeta_4\epsilon^2$ \cite{Anastasiou:2009kn}. For completeness we also
give the value of $C_{WL}^{(2)}=\frac{-\zeta_2}{2}$.

In what follows it will be useful to define $\Gamma_{cusp}^{(l)}=2 f_{0,WL}^{(l)}$and $\Gamma^{(l)}=2 f_{1,WL}^{(l)}/l$ and rewrite  \eqref{W_1in} as

\beqa\label{W_1}
\log W_1=
-\frac{1}{4}\sum_{l=1,2} a^l\big(\frac{\Gamma_{cusp}^{(l)}}{(l\epsilon)^2}
   +\frac{\Gamma^{(l)}}{l\epsilon}\big)\sum_{k'} (-\mu^2s_{k',k'+1}-i0)^{l\epsilon}    +a F_6(\mu^2,\epsilon,s_{k',j'})
      \nonumber\\
     +a^2f_{0,WL}^{(2)} F_6(\mu^2,2\epsilon,s_{k',j'})
+a^2f_{1,WL}^{(2)}\epsilon F_6(\mu^2,2\epsilon,s_{k',j'})
+R(\mu^2,\epsilon,s_{k',j'}) \nonumber\\
+C_{WL}^{(2)}-a^2\frac{f_{2,WL}^{(2)}}{8}
\sum_{k'}(-\mu^2 s_{k',k'+1})^{2\epsilon}
+\mathcal{O}(\epsilon)
\eeqa
In the last equation $F_6$ is the one loop finite part of the
Wilson loop. The second term of the second line of  \eqref{W_1}
is kept because it will give a finite contribution since, as discussed
 in Section 3, $F_6$ has a pole in $\epsilon$ for the loop of Figure 1a on which we will eventually focus. We have also kept the constants of the last line of
\eqref{W_1} although they will play no role in the rest of the paper.

For a generic configuration $F_6$ and $R$ are finite quantities
even in four dimensions. However, in what follows,
it is more convenient to think them as the functions that one would obtain
if one  was able to analytically evaluate the two loop integrals in $4-2\epsilon $ dimensions.
We want to keep $\epsilon\neq 0$ because this is the most effective way of regularising the crossing singularity.
For a generic loop one has the alternative of setting $\epsilon=0$ and evaluating the integrals in which case
one obtains a finite $\mu$-independent expression for $R$ which depends only on the cross ratios.

From \eqref{W_1} one can deduce the renormalised expression for $W_1^r$
 which reads:
 \begin{multline}
   \label{W_1^r}
   \log{W_1^r}=-\frac{1}{8}\sum_{k'}(\Gamma_{cusp}^{(1)}a+\Gamma_{cusp}^{(2)}a^2)\log^2(-\mu^2s_{k',k'+1}-i0)\\
   -\frac{1}{4}\sum_{k'}\Gamma^{(2)}a^2\log(-\mu^2s_{k',k'+1}-i0)
   +a i \pi \log(\mu^2s_{pq})+a^2 \frac{\Gamma_{cusp}^{(2)}}{2}
i\pi \log(\mu^2s_{pq})\\
+a \tilde{F}+a^2 \frac{\Gamma_{cusp}^{(2)}}{2}\tilde{F}+a^2 i \pi \frac{\Gamma^{(2)}}{2}
+R^r+C_{WL}^{(2)}-a^2 \frac{3 f_{2,WL}^{(2)}}{4}+\mathcal{O}(\alpha^3)
 \end{multline}
Some comments are in order. In \eqref{W_1^r} the $\mu$ dependent term involving the
invariant $s_{pq}$ come from the renormalisation of the one loop finite part involving a single gluon exchange between
the momenta $p$ and $q$ (see \eqref{intfin}) while $\tilde{F}$ denotes the $\mu$ independent one-loop
finite part of the loop in Figure 1a. As discussed previously, one should think of the two-loop finite remainder $R$
 as a function of $\epsilon$, $\mu$ and the kinematical variables which is finite and $\mu$
 independent at four dimensions for a generic amplitude, but may develop poles for the case we are considering.
$R^r$ is the renormalised value of $R$ after subtracting these $\epsilon$ poles, in case they are present.

We are almost in position to use plug the expressions for the renormalised loop functions in \eqref{RGinfty} in order to find the RGE that $R^r$ satisfies.
To this end, we divide the first equation of \eqref{RGinfty} by $W_1^r$
and by taking into account that the beta function of ${\cal N}=4$ is
zero, one can rewrite it as
\beqa\label{logW_1^r}
\mu\frac{\partial}{\partial \mu}\log{W_1^r}=i \pi \Gamma_1(g)\frac{W_2^r}{W_1^r}
-\frac{1}{2}\sum_{k'}\Gamma_{cusp}(g)\log(\mu^2(-s_{k',k'+1}-i0))-\tilde{\Gamma}(g).
\eeqa
By using the relations \eqref{W1W2} and \eqref{W_1^r}
we can write the first term on the right hand side of \eqref{logW_1^r} as
\beqa\label{W/W}
i \pi \Gamma_1(g)\frac{W_2^r}{W_1^r}=i \pi \big(\Gamma_1^{(1)}a+\Gamma_1^{(2)}a^2\big)
\Big(1-\frac{1}{8}\Gamma_{cusp}^{(1)}a \log^2(-\mu^2s_{pq}xy-i0)\nonumber\\
-\frac{1}{8}\Gamma_{cusp}^{(1)}a \log^2(-\mu^2s_{pq}(1-x)(1-y)-i0)
-\frac{1}{8}\Gamma_{cusp}^{(1)}a (\log^2(-\mu^2t_1)-\log^2(-\mu^2\frac{t_1}{x}))\nonumber\\
-\frac{1}{8}\Gamma_{cusp}^{(1)}a (\log^2(-\mu^2t_1)-\log^2(-\mu^2\frac{t_1}{y}))
-\frac{1}{8}\Gamma_{cusp}^{(1)}a (\log^2(-\mu^2t_2)-\log^2(-\mu^2\frac{t_2}{1-x}))\nonumber\\
-\frac{1}{8}\Gamma_{cusp}^{(1)}a (\log^2(-\mu^2t_2)-\log^2(-\mu^2\frac{t_2}{1-y}))
-a i \pi \log(\mu^2s_{pq})+aF_1+aF_2-a\tilde{F}\Big)
\eeqa

We should mention that in the right hand side of \eqref{W/W} one should keep terms up to order $a^2$, which means that one should use the order $a$ expressions for $W_1^r$ and $W_2^r$ since their ratio is multiplied by $\Gamma_1$ which is
already of order $a$.

We are now in position to plug \eqref{W_1^r} and \eqref{W/W} into \eqref{logW_1^r} and derive the equation that $R^r$ satisfies:
\beqa\label{dR^r}
\mu\frac{\partial}{\partial \mu}R^r=-\frac{i \pi}{8}\Gamma_{cusp}^{(1)}\Gamma_1^{(1)}a^2
   \Big(\log^2(-\mu^2s_{pq}xy-i0)+ \log^2(-\mu^2s_{pq}(1-x)(1-y)-i0)\nonumber\\
   +\log^2(-\mu^2t_1)-\log^2(-\mu^2\frac{t_1}{x})+\log^2(-\mu^2t_1)-\log^2(-\mu^2\frac{t_1}{y})
+\log^2(-\mu^2t_2)\nonumber\\
   -\log^2(-\mu^2\frac{t_2}{1-x})+\log^2(-\mu^2t_2)-\log^2(-\mu^2\frac{t_2}{1-y})
   \Big)+\pi^2 a^2\Gamma_1^{(1)} \log(\mu^2 s_{pq})\nonumber\\
    +i \pi \Gamma_1^{(1)}a^2(F_1+F_2-\tilde{F})+i \pi\Gamma_1^{(2)}a^2
         - \tilde{\Gamma}(g) +\frac{1}{2}\sum_{k'}\Gamma^{(2)}a^2-i \pi a^2
\Gamma_{cusp}^{(2)}
\eeqa
The two last terms in \eqref{dR^r} originate from the differentiation of the first and third term in the second line of \eqref{W_1^r} respectively.
Using the values for $\Gamma_1^{(1)}=\Gamma_{cusp}^{(1)}=2$ and
$\Gamma_{cusp}^{(2)}=-2 \zeta_2$
this equation can be easily integrated to give:
\beqa\label{R^r}
R^r=-\frac{i \pi}{12}a^2\Big(\log^3(-\mu^2s_{pq}xy)+\log^3(-\mu^2s_{pq}(1-x)(1-y))
   +2\log^3(-\mu^2t_1)&&\nonumber\\
   -\log^3(-\mu^2\frac{t_1}{x})-\log^3(-\mu^2\frac{t_1}{y})
   +2\log^3(-\mu^2t_2)-\log^3(-\mu^2\frac{t_2}{1-x})-\log^3(-\mu^2\frac{t_2}{1-y})
   \Big)&&\nonumber\\
   +\frac{1}{2}\pi^2a^2 \log^2(\mu^2 s_{pq})
      +\frac{1}{2}\Big(i \pi 2 a^2(F_1+F_2-\tilde{F})+i \pi\Gamma_1^{(2)}a^2
          +\frac{1}{2}\sum_{k'}\Gamma^{(2)}a^2- \tilde{\Gamma}(g)+i \pi a^2 2\zeta_2\Big)&&\nonumber\\
\log(-\mu^2s_{pq}-i0)&&
\eeqa
which implies after some algebra
\beqa\label{R_cfin}
R=-\frac{i \pi a^2}{8\epsilon^3}(-\mu^2s_{pq})^{2\epsilon}
  -\frac{i \pi a^2}{4\epsilon^2}(-\mu^2s_{pq})^{2\epsilon}\log(xy(1-x)(1-y))
  +\frac{ \pi^2 a^2}{4\epsilon^2}(\mu^2s_{pq})^{2\epsilon}\nonumber\\
  -\frac{i \pi a^2}{8\epsilon}(-\mu^2s_{pq})^{2\epsilon}Y
  +\frac{a^2}{2\epsilon}(-\mu^2s_{pq})^{2\epsilon}T+C',
\eeqa
where
\beqa\label{Q}
Y=\Big(\log^2(xy)+\log^2((1-x)(1-y))
  +2\log^2\frac{t_1}{s_{pq}}-\log^2\frac{t_1}{s_{pq}x}
  -\log^2\frac{t_1}{s_{pq}y}\nonumber\\
  +2\log^2\frac{t_2}{s_{pq}}-\log^2\frac{t_2}{s_{pq}(1-x)}-\log^2\frac{t_2}{s_{pq}(1-y)}\Big)\nonumber\\
T=i \pi(F_1+F_2-\tilde{F})+\frac{i \pi}{2}\Gamma_1^{(2)}-\frac{1}{4}\sum_{k'}\Gamma^{(2)}-\frac{1}{2}\tilde{\Gamma}^{(2)}+i \pi \zeta_2.
\eeqa
One can check that the finite part obtained from \eqref{R_cfin} after throwing
away its poles is equal to the result of \eqref{R^r} up to
$\mu$ independent terms, which the RGE do not control, anyway.
Of course, one has to do a bit of algebra to bring the logs of \eqref{R^r}
in the form $\log(-\mu^2s_{pq})$.

From \eqref{R_cfin} it is evident that the unrenormalised expression for $R$  contains
poles in \\
$1/\epsilon^k (-\mu^2 s_{pq})^{2 \epsilon}$, where $k \leq 3$.
These poles originate from Feynman diagrams which have two gluons, at least,
attached to the intersecting legs $p$ and $q$. Notice, however, that not all diagrams with
two gluons attached to the intersecting legs give such poles. For instance, the
non-abelian part of the diagram where two gluons are exchanged, one between $q$ and $p_2$
and the other between $q$ and $p_3$ contains no such pole.
From \eqref{R_cfin} we can read the leading singularity of R to be
$-\frac{i \pi a^2}{8\epsilon^3}$.
At first sight, this $1/\epsilon^3$ pole seems peculiar since as discussed
in \cite{Drummond:2007cf} all the terms with poles of order higher than two cancel in the final result for the Wilson loop. However, this should  not be
the case for an intersecting loop.
Its leading $1/\epsilon^3$ singularity comes from diagrams where all the gluons
are exchanged between the momenta that cross.
The subleading poles can also originate from diagrams where one or more gluons
are not attached to the crossing gluons.

As discussed at the end of section 3, one can regularise the crossing configuration
either by dimensional regularisation or by allowing a small distance between the
intersecting gluons. What is $1/\epsilon$ in the former approach becomes $-\log(1-u_2)$
in the latter. This implies that the leading behaviour of the finite part $R$
in  the case where the two lines almost cross is
\beqa\label{RR}
R\sim i \pi \log^3(1-u_2).
\eeqa
Although what concern us here is the leading singularity \eqref{R^r} points
that $R$ will also have subleading terms which behave like $\log^2(1-u_2)$ and
$\log(1-u_2)$.
Thus, we conclude that the finite remainder function develops a divergent imaginary part
as one approaches the crossing configuration
\footnote{For the case of almost intersecting lines, we have verified that
the non-abelian part of the diagram where 2 gluons are exchanged between $p$ and $q$  is indeed proportional to $-i \pi \log^3(-c)$ when approaching the crossing configuration.}.
Finally, let us point out that our approach of studying the loop behaviour of the MHV remainder function by considering self-crossing configurations was further developed in a series of nice papers  \cite{Dorn}.

\sect{Discussion}
In this section, we discuss some of the implications of the results
obtained in the two previous sections.

The one-loop n-gluon MHV scattering amplitude in $N=4$ SYM was firstly
evaluated in \cite{Bern:1994zx} using unitarity methods, as well as information
coming from taking appropriate collinear limits.
To simplify the calculation of the integrals, the authors of \cite{Bern:1994zx}
restricted the calculation in the kinematical region where all the Lorentz
invariants are negative (Euclidean region). Their result was that the
helicity-blind part of the amplitude is a sum over all different
two-mass easy box functions $F^{2me}$, all having coefficient equal to one.
Namely,
\beqa\label{M_1}
M_1^{(1)}=\sum_{p,q}F^{2me}(p,q,P,Q),
\eeqa
where $F^{2me}$ is given by
\begin{multline}
\label{2me}
F^{2me}(p,q,P,Q)=-\frac{1}{2\epsilon^2}\Big(
     (\frac{-s}{\mu^2_{IR}})^{-\epsilon_{IR}}+
     (\frac{-t}{\mu^2_{IR}})^{-\epsilon_{IR}}-
     (\frac{-P^2}{\mu^2_{IR}})^{-\epsilon_{IR}}-
     (\frac{-Q^2}{\mu^2_{IR}})^{-\epsilon_{IR}}\Big)\\
    +\frac{1}{2}\big(\frac{1}{2}\log^2(\frac{s}{t}) +Li_2(1-\frac{P^2}{s})+
      Li_2(1-\frac{P^2}{t})+Li_2(1-\frac{Q^2}{s})+Li_2(1-\frac{Q^2}{t})-
      Li_2(1-\frac{P^2Q^2}{st})\big)
\end{multline}
and $c$ is given by \eqref{c}.
If ones wishes to write down the amplitude in the physical region, one
should analytically continue the expression \eqref{2me}. Usually, this
analytic continuation to positive values of the kinematic variables is
 achieved by applying the replacement
$(kinematic\, invariant)\rightarrow (kinematic\, invariant)+i 0^+$.
As was noticed in \cite{Binoth:1999sp} this naive continuation works fine for
most of the terms in  \eqref{2me}. However, the last dilogarithm of
\eqref{2me} needs special care. One has to be careful to not cross its cut
which extends from one to infinity on the real axis.
The correct analytic continuation for this term is achieved
by making the replacement
\beqa\label{continuation}
&&Li_2(1-\frac{P^2Q^2}{st})\rightarrow
Li_2(1-\frac{P^2+i0}{s+i0}\frac{Q^2+i0}{t+i0})+\\
&& \Big(\log(\frac{P^2+i0}{s+i0}\frac{Q^2+i0}{t+i0})
    -\log(\frac{P^2+i0}{s+i0})-\log(\frac{Q^2+i0}{t+i0})\Big)
       \log(1-\frac{P^2+i0}{s+i0}\frac{Q^2+i0}{t+i0})\nonumber.
\eeqa
When supplemented with the second line of \eqref{continuation},
\eqref{2me} provides an expression for the two-mass easy box
function which is valid for all kinematical regimes.
A second representation of the  two-mass easy box, which is also
valid for all kinematical regimes, was obtained in \cite{Duplancic:2000sk},\cite{Brandhuber:2005kd}.
The  expression for its  finite part is given in \eqref{e=0}.
The equivalence of the latter with \eqref{2me}
,properly analytically continued, was shown numerically
in \cite{Duplancic:2000sk}.
It is elementary to check that it is precisely the  non-trivial analytic
continuation term appearing in \eqref{continuation}
which gives the $- \pi i \log(1-u_2)$ of \eqref{e=0appfin} (see\eqref{cu}).

The contribution to the amplitude from the first term of \eqref{cu},
which is not a function of the invariant cross ratios $u_i,i=1,2,3$ has to
cancel against an opposite contribution coming from the single gluon exchanges
between non-adjacent legs. The reason for this is that
the amplitude obeys the anomalous conformal Ward identity
\cite{Drummond:2007au}
-we remind the reader that we are now considering a case where
 $u_2$ is almost but not exactly one-
\beqa\label{Ward}
K^{\nu}F_n^{(WL)}\equiv\sum_i^{n} (2x_i^{\nu} x_i\cdot\partial_i
   -x_i^2\partial_i^{\nu})F_n^{(WL)}=\frac{1}{2}\Gamma_{cusp}
       \sum_i^{n}x_{i,i+1}^{\nu}\log(\frac{x_{i,i+2}^{2}}{x_{i-1,i+1}^{2}})
\eeqa
in any kinematical region.
Furthermore, in \cite{Drummond:2007cf} it was checked that the one-loop n-gluon amplitude
satisfies \eqref{Ward} when all the invariants lie in the Euclidean region.
This implies that any terms coming from the non-trivial
continuation discussed above should depend only on the conformal cross ratios.
Since all such terms come from the last dilog of \eqref{2me}
they should be proportional to $\log(1-u_2)$, see \eqref{continuation}.
This means that the non-conformal part of \eqref{e=0appfin}
(see also\eqref{cu}) should cancel in the whole amplitude.
We have checked, that this is, indeed, the case.
Furthermore notice that, if one was computing numerically the one loop amplitude and
had restricted himself to the Euclidean region, he would have obtained real
values for all the integrals. In other words, he would have completely missed
 the second line of \eqref{continuation} and the fact that $F^{2me}$
has a branch cut related to the last dilog of \eqref{2me}.

We now proceed to some comments regarding the behaviour of the two-loop
remainder function $R$. In a nice paper \cite{Anastasiou:2009kn}, this function was
numerically evaluated for  a wide range of the kinematic variables.
One of the kinematic points the authors of \cite{Anastasiou:2009kn} considered was
the point with conformal ratios $u_1=u_2=u_3=1$ for which the remainder
function was found to be $R=-2.70...$.
This result is, apparently, in contradiction with \eqref{R_cfin} (see also \eqref{RR})
and the discussion below it which predicts an infinite imaginary part
for $R$. However, one should not forget that the calculation of \cite{Anastasiou:2009kn}
was performed in the Euclidean region,
whereas our result holds in the physical region.
This suggests that, as in the one-loop case,
 the behaviour of the finite remainder $R$ dictated by
\eqref{R_cfin} is due to a non-trivial analytical continuation analogous to the
one needed for the one-loop finite part and as such
it could never be seen in the
numerical calculations of \cite{Anastasiou:2009kn},\cite{Drummond:2008aq}.

Since no analytic expression for
the finite remainder function $R$ is known, let us see
if we can get any information about it.
The discussion at the end of Section 4 shows that near the crossing
configuration , that is as $u_2\rightarrow1$ and  $ u_1\approx u_3$,
the remainder function $R$ behaves like $R\sim i \pi c_1 \log^3(1-u_2)+...$, where $c_1$ is a
rational constant.
The simplest function
reproducing
the aforementioned leading behaviour of $R$ when  analytically continued, in a similar fashion to \eqref{continuation}, is
$Li_{4}(1-u_2)$.
In order to avoid crossing the polylog's cut, one has to add
additional terms proportional to the discontinuity
 of the function along the cut.
 Taking into account that $\lim_{\epsilon\rightarrow 0_+}(Li_n(x+i \epsilon)-Li_n(x-i \epsilon))=\frac{2 \pi i \log^{n-1}(x)}{\Gamma(n)}$ it is evident that the limiting behaviour of $R$ is consistent with the analytic continuation of $Li_{4}(1-u_2)$.
In any case, we should stress that the choice above is by no means unique, since the relation $R\sim i \pi \log^3(1-u_2)$
holds only as we asymptotically approach the crossing.
However, this behaviour indicates that near the crossing $R$ includes a function of $1-u_2$ which has a branch cut
for negative values of $u_2$. Thus, it is conceivable that $R$ has a cut along the negative $u_2$ axis when $u_1\approx u_3$.

It is clearly desirable to obtain an analytic expression for the remainder function. The positions where its cuts are situated as well as the discontinuities along these cuts can play a crucial role to its determination.
Thus, it is evident that the analytic properties of $R$ are very important.
An interesting discussion of the analytic structure of the Wilson loop
diagrams was presented in \cite{Gorsky:2009nv}.

We close this note by making two remarks. The first
concerns the transcendentality of $R$. All the terms in \eqref{R_cfin} have the same transcendentality 4,
in agreement with the transcendentality principle, provided that $\Gamma_1^{(2)}$ and $\tilde{\Gamma}^{(2)}$
have transcendentality 2 and 3, respectively.
The second  concerns the fact that it is not possible to construct crossing
 configurations for the scattering of 4 and 5 gluons. This is in agreement with
the fact that a non-trivial remainder function $R$ starts appearing from 6 gluons on.\\

\
\vspace{1cm}

\noindent {\large {\bf Acknowledgments}}

\vspace{3mm}

\noindent
We wish to thank Andreas Brandhuber, Valeria Gili, Paul Heslop, Rodolfo Russo
and especially Gabriele Travaglini for useful discussions and comments.
G.G. was partly supported by
STFC through a Postdoctoral Fellowship.


\end{document}